\documentclass[a4paper]{jpconf}
\usepackage{graphicx}

\begin{document}


\title{Photoassociative molecular spectroscopy for atomic radiative lifetimes.\\}

\author{Nadia Bouloufa, Anne Crubellier, Olivier Dulieu}

\address{Laboratoire Aim\'e Cotton, UPR3321, CNRS, b\^at 505, Univ Paris-Sud , 91400 Orsay,
France}

\ead{anne.crubellier@lac.u-psud.fr}

\date{\today}

\begin{abstract}
When the atoms of a dimer remain most of the time very far apart, in so-called long-range
molecular
states, their mutual interaction is ruled by plain atomic properties. The high-resolution
spectroscopic
study of some molecular excited states populated by photoassociation of cold atoms
(photoassociative
spectroscopy) gives a good illustration of this property. It provides accurate determinations
of atomic
radiative lifetimes, which are known to be sensitive tests for atomic calculations. A number
of such
analyses has been performed up to now, for all stable alkali atoms and for some other atomic
species (Ca,
Sr, Yb). A systematic review of these determinations is attempted here, with special attention
paid to
accuracy issues.
\end{abstract}


\section{Introduction}
The precise knowledge of atomic radiative lifetimes is a prerequisite for many problems of
modern quantum
physics. First of all, their measurement often represents an accurate check for {\it ab
initio}
calculations of atomic structure \cite{jonsson1996,porsev2001,fischer2003}: indeed, their
computed values
are much more sensitive to the details of the electronic wave functions than the computed
total energy.
The reliability of such calculations is essential when atoms are used for the investigation of
fundamental
problems, like the search for parity non-conservation effects which are predicted by theories
beyond the
Standard Model. Up to now, the strongest constraint on the magnitude of such effects is
provided by
experiments with cesium \cite{bennett1999,guena2005,derevianko2007} for parity violation, or
with thallium
atoms \cite{regan2002} for time-reversal symmetry violation which would manifest itself by a
permanent
electric dipole moment of the electron.

Among all atomic species, the alkali atoms are systems of choice for comparative experimental
and
theoretical studies: their main resonant transition $^2S\rightarrow$$^2P$ lies in the optical
domain,
while calculations are facilitated by their simple electronic structure. At the 5$^{th}$ ASOS
conference
in Meudon (France) in 1995, U. Volz and H. Schmoranzer presented a review on measurements and
calculations
of alkali atom radiative lifetimes, as well as a series of updated precision measurements of
the radiative
lifetimes for Li, Na, K, and Rb atoms using beam-gas-laser spectroscopy \cite{volz1996}. In
particular,
they solved a long-standing discrepancy between ab-initio calculations and measurements for
lithium and
sodium. In their paper, they refered to the first measurement of a radiative lifetime using
the emerging
technique of photoassociative spectroscopy (PAS) on ultracold lithium atoms
\cite{mcalexander1995}. Their
result was in good agreement with the PAS one, while the error bar derived with the PAS
technique was
claimed to be four times smaller than their own.

More than a decade later, and motivated by our recent work on the radiative lifetime of atomic
cesium
\cite{bouloufa2007}, we take the opportunity offered by the edition of the special issue of
Physica
Scripta for the 9$^{th}$ ASOS conference, to review the status and the accuracy of the PAS
approach for
determining the radiative lifetimes of alkali atoms, and more generally of atomic species which
are
nowadays laser-cooled and trapped. In Section \ref{sec:pas}, we briefly describe the
photoassociation
process which allows the investigation of the long-range interaction within a pair of cold
atoms,
focussing on its link with the energy-level spacing of the atom pair through the so-called
LeRoy-Bernstein
law, and with the atomic radiative lifetime. Next (Section \ref{sec:0g-}) we focus on a
specific class of
electronic states of the atom pair - the so-called "pure long-range" states - which are
particularly
relevant for the extraction of Na \cite{jones1996}, K \cite{wang1997b}, Rb
\cite{freeland2001,gutterres2002} and Cs \cite{amiot2002,bouloufa2007} radiative lifetimes
from PAS of
these unusual molecular states. The case of molecular states which are not of the pure
long-range class is
described in Section \ref{sec:longrange}. Two different methods have been used to derive
accurate
radiative atomic lifetimes from PAS performed in cold samples: a direct application of the
LeRoy-Bernstein
asymptotic law has been used in the cases of Sr \cite{yasuda2006} and Yb \cite{takasu2004}; a
global fit
of the full molecular potential at both short and large distances has been performed to
extract the
radiative lifetimes for Li \cite{mcalexander1996}, Ca \cite{vogt2007} and Sr \cite{nagel2005}
atoms. The
last section is devoted to the sensitive issue of the evaluation of the accuracy of the
results for the
various species obtained by PAS, which is generally claimed to be better than the one obtained
from
standard atomic physics techniques.

\section{Photoassociation of cold atoms, photoassociation spectroscopy
	and long-range interactions}				           													
\label{sec:pas}

\subsection{Basics of photoassociation process}																	 
\label{sec:PA}																																 

\begin{figure}[h]
\includegraphics[width=30pc]{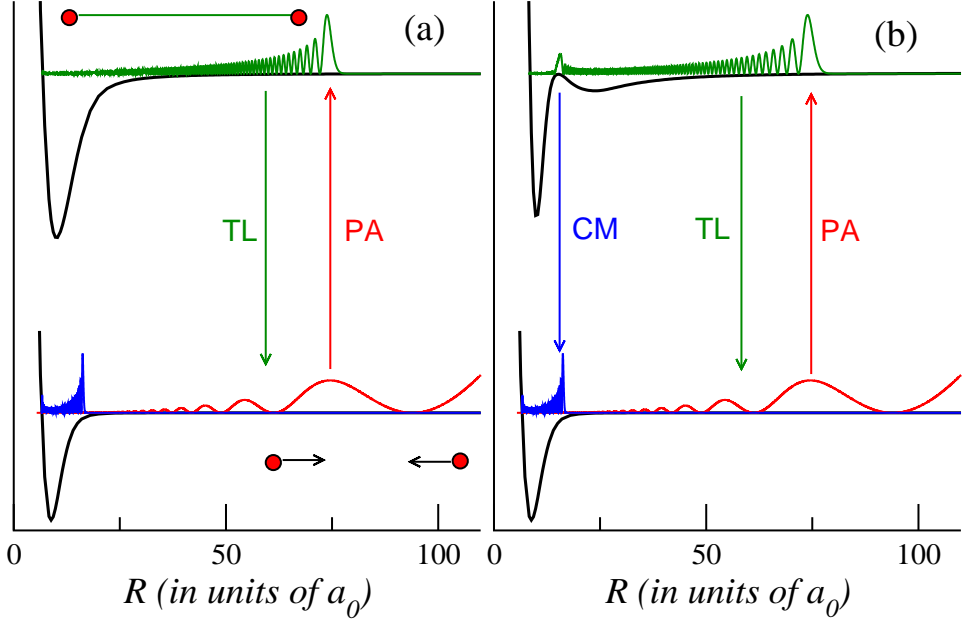}
\caption{\footnotesize General scheme of the photoassociation of a pair of cold atoms in their
$S$ ground
state. In panel (a), the atoms interact through a molecular potential curve, with a van der
Waals
character at large $R$, {\it i.e.}, varying as $R^{-6}$. As their relative initial kinetic
energy is very
small, the radial wave function has a large amplitude at large distances, so that in a
classical vision,
the atom pair likely absorbs a photon with a frequency red-detuned from the atomic
$S\rightarrow P$
transition frequency (arrow labelled PA). Bound levels with a small binding energy and a large
spatial
extension are efficiently populated. Most of the relative vibrational motion then takes place
at large
$R$, where the excited interaction potential behaves as $R^{-3}$. This bound level can either
decay back
to a pair of free atoms which escape from the cold atom trap (arrow TL, for trap loss), or to
a bound
level of the molecular ground state, creating then a cold molecule (arrow CM, for cold
molecule). Panel
(b) illustrates the same process for an excited double-well potential like, for instance, in
cesium
photoassociation.}
\label{fig:pa}
\end{figure}

Photoassociation (PA) in atomic vapors is a well-known process
\cite{pichler1983,jones1993,marvet1995,ban1999}: a pair of atoms (M,M') absorbs a photon of
suitable
energy $h\nu_{PA}$, generally red-detuned from the energy of an atomic transition $h\nu_{0}$,
to create
molecules in rovibrational levels of an excited electronic state according to the reaction:
$M+M'+h\nu_{PA} \rightarrow (MM')^*$. At room temperature, the PA process is not selective in
the final
state, due to the width of the Maxwell-Boltzmann kinetic energy distribution of the atoms,
which is most
often larger than the energy gap between consecutive molecular levels. Shortly after the first
experimental observations of laser-cooling of atoms, Thorsheim et al \cite{thorsheim1987}
proposed to
perform photoassociation of ultracold atoms: as the width of the kinetic energy distribution
of the cold
atoms is comparable or even smaller than the natural width of the excited state, the
free-bound PA process
allows a selective excitation of the initial atom pair into an excited molecular state, just
like a
bound-bound process would do. Then it is possible to reach rovibrational levels with large
spatial
extension and with small binding energy (Figure \ref{fig:pa}). PA has been observed a few
years later on
cold sodium \cite{lett1993} and rubidium \cite{miller1993} samples, and became soon a
fantastic tool for
high-resolution molecular spectroscopy, referred to as photoassociation spectroscopy (PAS)
\cite{stwalley1999}. The recent review by Jones et al \cite{jones2006} yields a comprehensive
study of
PAS, and addresses one of the main
issues of PA which is particularly relevant for the present paper: "PA favors the study of
physicists'
molecules, {\it i.e.}, molecules whose properties can be related (with high precision) to the
properties
of the constituent atoms.".
\subsection{Long-range resonant dipole behavior}								
\label{sec:dipole-dipole}											

All the experiments reviewed here follow the general scheme of Figure \ref{fig:pa}. A pair of
identical
atoms in their ground state (either $ns$~$^2$S$_{1/2}$ or $ns^2$~$^1$S$_0$ for alkali or
alkaline-earth-like elements, respectively) is excited into a molecular state connected to the
lowest
($ns$~$^2$S$_{1/2}$+$np$~$^2$P$_{1/2,3/2}$) or ($ns^2$~$^1$S$_0$+$nsnp$~$^1$P$^0_1$)
dissociation limit.
In all these cases, the asymptotic interaction between the atoms is dominated by the resonant
dipole
interaction and is approximated by
\begin{eqnarray}
V(R)=D-\frac{C_3}{R^3},
\label{eq:potdipdip}
\end{eqnarray}
where $D$ is the dissociation energy of the interaction potential. The coefficient $C_3$
depends on the
relative orientation of the atomic dipoles with respect to the molecular axis  and is thus
different for a
$\Sigma$ or a $\Pi$ molecular state. It corresponds to the exchange of excitation between the
two atoms
and it is simply related to a characteristic of the atom, the atomic dipole matrix element.
One has
\begin{eqnarray}
C_3^{\Sigma}=-2C_3^{\Pi}=-2\frac{\mathcal{R}^2}{3},
\label{eq:c3em}
\end{eqnarray}
where $\mathcal{R}=<ns|r|np>$ is the radial integral of the dipole length operator between the
atomic $ns$
and $np$ orbitals. The $C_3$ coefficients are therefore related to the atomic radiative
lifetime of the
$P$ level since
\begin{eqnarray}
\tau=\frac{9\hbar \lambda ^3}{32 \pi ^3 \mathcal{R}^2},
\label{eq:tauem}
\end{eqnarray}
where $\lambda$ is the wavelength of the considered $S\rightarrow P$ atomic transition.

Equation (\ref{eq:potdipdip}) is always an approximation, only valid for very large
interatomic distances.
Several other effects are likely to contribute to the interaction energy:

\begin{itemize}
\item the following terms $C_n/R^n$ with n=6, 8, 10, .... of the multipole expansion: they
account for
polarization and dispersion forces;
\item the spin-orbit interaction: it has to be taken into account for alkali atoms, but not
for
alkaline-earth and alkaline-earth-like ones, due to the choice of the excited state;
\item the hyperfine structure of the ground state: for alkali atoms, the hyperfine splitting
of the
excited atomic state increases with atomic mass; it is almost negligible at the experimental
precision of
PAS for lithium but not for cesium; the lack of the hyperfine structure for the chosen
isotopes of the
other elements considered here, Ca, Sr and Yb, simplifies the analysis of the experiments;
\item the rotation of the molecule: it is often absent in the initial colliding state, when
the
temperature of the atom cloud is low enough ($s$-wave collisions); however it has always to be
accounted
for in the excited molecular state.
The rotational energy $E^{rot}(v,J)$ of a rovibrational level $(v,J)$ can be estimated by the
diagonal
part of the rotational
Hamiltonian
\begin{eqnarray}
H^{rot}=\frac{\mathbf{J}^2-2\mathbf{J}.\mathbf{j}+\mathbf{j}^2}{2\mu R^2},
\label{eq:Hrot}
\end{eqnarray}
where $\mathbf{J}$ and $\mathbf{j}$ are, respectively, the total angular momentum and the
total electronic
angular momentum of the considered dimer and $\mu$ its reduced mass;
\item the overlap between the two electron clouds: it is manifested by an exchange energy
which vanishes
exponentially with increasing atomic distances \cite{evseev1978,marinescu1996}. For the
excited states of
the alkali dimers that we consider here, it depends essentially on a single parameter which is
the product
of the amplitudes of the atomic $ns$ and $np$ wavefunctions. The derivation of the relevant
expressions of
the $\Sigma$ and $\Pi$ asymptotic exchange terms, $V_{exch}^{\Sigma,\Pi}$, is recalled in Ref
\cite{amiot2002};
\item the retardation effect, related to the Casimir-Polder effect in London-van der Waals
interaction
\cite{casimir1948}: when accounting for the finite velocity of light, the long-range
interaction between
the atoms is modified. This effect is usually small but clearly noticeable in several
experiments reviewed
here. Following Refs \cite{mclone1965,meath1968}, it can be accounted for simply, by
multiplying the $C_3$
coefficient by a correcting term which is different for $\Pi$ and $\Sigma$ states:
\begin{eqnarray}
f^{\Pi}(R)&=&\cos\big(\frac{2\pi R}{\lambda}\big)+\big(\frac{2\pi
R}{\lambda}\big)\sin\big(\frac{2\pi
R}{\lambda}\big)-
\big(\frac{2\pi R}{\lambda}\big)^2\cos\big(\frac{2\pi R}{\lambda}\big)\\
f^{\Sigma}(R)&=&\cos\big(\frac{2\pi R}{\lambda}\big)+\big(\frac{2\pi
R}{\lambda}\big)\sin\big(\frac{2\pi
R}{\lambda}\big);
\label{eq:retard}
\end{eqnarray}
\item the intra-atomic relativistic effects: they tend to contract differently the atomic
$p_{1/2}$ and
$p_{3/2}$ orbitals \cite{aymar2006}. It is possible to account for these effects through a
small parameter
$\epsilon$, which characterizes the ratio between the dipole matrix elements corresponding to
the two $p$
orbitals
\begin{eqnarray}
\frac{<ns|r|np_{3/2}>^2}{<ns|r|np_{1/2}>^2}=\frac{2\tau_{1/2}}{\tau_{3/2}}\Big(\frac{\lambda_{
3/2}}{\lambda_{1/2}}\Big)^3=\frac{2}{(1+\epsilon)^2}.
\label{eq:rel}
\end{eqnarray}
More details can be found in rubidium \cite{gutterres2002} and cesium
\cite{amiot2002,bouloufa2007}
studies;
\item the non-adiabatic terms negelected in the Born-Oppenheimer approximation, which assumes
fixed nuclei
of infinite mass: it is possible \cite{bunker1968} to release this assumption while still
maintaining the
decoupling of nuclear and electronic motions, by considering the diagonal corrections for the
motion of
the nuclei. The description of these terms can be found in Ref \cite{mcalexander1996};
\item the specific shift and broadening of the PA lines due to the temperature of the atomic
cloud: these
effects depend also on the shapes of the ground and excited molecular potentials. They are
discussed in
great detail in the calcium study \cite{vogt2007};
\item the so-called predissociation process: it is due to the interaction of the molecular
state with the
continuum of a neighboring one and it gives rise to line broadening. This problem arose in
particular in
the ytterbium study \cite{takasu2004}.
\end{itemize}

\subsection{LeRoy-Bernstein law and energy spacing of high-lying 						
	vibrational states}											

\label{sec:LRB}

Assuming that the asymptotic form of the molecular potential is written as
\begin{eqnarray}
V(R)=D-\frac{C_n}{R^n},
\label{eq:potasympt}
\end{eqnarray}
one can show, using a semi-classical Wigner-Kramers-Brillouin (WKB) description of the
vibrational
wavefunction, that the energy $E_v$ of a molecular vibrational level is related to its
vibrational quantum
number $v$ through the so-called LeRoy-Bernstein law \cite{leroy1970}:
\begin{eqnarray}
(D-E_v)=[X_n(v_D-v)]^{(2n/n-2)} \cr \cr
X_n=\hbar \sqrt{\frac{\pi}{2\mu}}~\frac{(n-2)\Gamma(1+1/n)}{\Gamma(1/2+1/n)}~C_n^{-1/n}.
\label{eq:lrb}
\end{eqnarray}
In Equation (\ref{eq:lrb}), $\Gamma(z)$ is an Euler Gamma function and $v_D$ is a non-integer
number (the
non-integer value that $v$ would take at the dissociation limit $D$) which is related to the
phase
accumulated by the wavefunction over its whole spatial extension, including the short-range
part of the
potential. In the case of a $R^{-3}$ potential, Equation (\ref{eq:lrb}) becomes
\begin{eqnarray}
(D-E_v)=X_0(v_D-v)^{6},
\label{eq:lrbbis}
\end{eqnarray}
with
\begin{eqnarray}
X_0=(X_3)^6=\Big[\frac{\hbar^2}{2\mu}\Big]^3\Big[\frac{\Gamma(4/3)}{\Gamma(5/6)}\Big]^6\frac{\
pi^3}{(C_3)^
2}.
\label{eq:x0}
\end{eqnarray}
As the total number of vibrational levels of the potential is most often not known, it is
convenient,
instead of labeling the levels by a number $v$ starting from the $v=0$ lowest vibrational
level, to
characterize them by $v^*$ starting from the $v^*=1$ uppermost one. One has then
\begin{eqnarray}
(D-E_v)=X_0(v^*-v^*_D)^{6},
\label{eq:lrbter}
\end{eqnarray}
with $v_D$ reduced to its fractional part $0\le v^*_D < 1$. Assuming that the density of
levels allows
one to introduce a continuous derivative with respect to $v$, the energy spacing between
consecutive
vibrational levels can be written as
\begin{eqnarray}
\frac{dE}{dv}=-\frac{2n}{n-2}(X_n)^{2n/(n-2)}(v_D-v)^{(n+2)/(n-2)}
\label{eq:energyspacing}
\end{eqnarray}
or, in the $n=3$ case,
\begin{eqnarray}
\frac{dE}{dv^*}=-6X_0(v^*-v^*_D)^5
\label{eq:energyspacingx0}
\end{eqnarray}
and appears thus as depending directly on $X_0$, {\it i.e.} on $C_{3}$ (Equation
(\ref{eq:x0})): one
clearly sees how the analysis of PAS data allows one to extract the atomic lifetime from the
energy
spacing of high-lying levels.

\subsection{Pure long-range states: molecules as atom pairs}  							
\label{pureLR}

For Na$_2$, K$_2$, Rb$_2$ and Cs$_2$ dimers, the molecular state of $0_g^-$ symmetry
converging toward the
first ($ns+np_{3/2}$) limit is a "pure long-range" state according to the definition of Ref
\cite{stwalley1978}. It has a double-well potential, for which the most external well is
shallow and
entirely located at unusually large internuclear separations.
Each atom keeps somehow its identity and the molecule looks more like a pair of atoms. The
electron cloud
overlap is unimportant, so that the potential is completely determined by long-range
interatomic forces
and atomic spin-orbit splitting, and can thus be calculated with high precision. We emphasize
here that
this situation is very different from the one encountered with usual molecular potentials, for
which the
knowledge of the inner part of the potential curve most often relies on quantum chemistry
calculations,
which never reach such a precision.

The dominant long-range interactions are the $R^{-3}$ resonant dipole interaction and the
spin-orbit
coupling \cite{dashevskaya1969}. For the $0_g^-$ potentials, the simple analytical model
introduced in
\cite{movre1977} is the basis of the analysis. Two $0_g^-$ potentials arise from a mixing of
two Hund's
case (a) states, a repulsive $^3\Pi_g$ state and an attractive $^3\Sigma_g^+$ one. The two
adiabatic
$0_g^-$ potentials are obtained by diagonalizing the matrix of the interaction in this basis
of states,
\begin{eqnarray}
&& \hspace{1.75cm} \Pi \hspace{1.75cm} \Sigma \nonumber \\
V_{MP}(R)&=&\left( \begin{array}{cc}
V^{\Pi}(R)-\frac{2\Delta}{3} & \frac{\sqrt{2}\Delta}{3} \\
\frac{\sqrt{2}\Delta}{3} & V^{\Sigma}(R)-\frac{\Delta}{3}
\end{array} \right) \begin{array}{ll} \Pi \\ \Sigma \end{array},
\label{eq:MPplus}
\end{eqnarray}
with
\begin{eqnarray}
V^{\Pi/\Sigma}(R)=-\frac{C_3^{\Pi/\Sigma}}{R^3}f^{\Pi/\Sigma}(R)-\frac{C_6^{\Pi/\Sigma}}{R^6}-
\frac{C_8^{\
Pi/\Sigma}}{R^8}+V_{exch}^{\Pi/\Sigma},
\label{eq:potentials}
\end{eqnarray}
where $\Delta$ is the atomic spin-orbit splitting and the zero of energy is taken at the
($ns+np_{3/2}$)
asymptote. The $0_g^-$ potential in which we are interested is the upper adiabatic potential,
which
converges to the ($ns+np_{3/2}$) limit. As this state is a mixture with $R$-varying weights of
two states
having different coefficients of the $R^{-3}$ term, it is clear that one has an "effective"
$C_3$ which
varies with $R$ and that the validity of the LeRoy-Bernstein law is limited by the mixing of
states.
The knowledge of the eigenvectors of the matrix of Eq.(\ref{eq:MPplus}) allows one to evaluate
the
corrections which involve the $R$-dependent mixing of states, like the $\mathbf{j}^2$ term of
Equation
(\ref{eq:Hrot}), or which contain $R$-derivatives, like the non Born-Oppenheimer corrections.
The
determination of the parameters entering in the matrix of Eq.(\ref{eq:MPplus}), where it is
possible to
add all the effects previously described (see for instance \cite {gutterres2002}), yields an
analytical
expression for the $0_g^-$ potential.

\section{Determination of the radiative lifetime of N\lowercase{a}, K, R\lowercase{b},
C\lowercase{s}
atoms
from pure long-range state analysis}
\label{sec:0g-}

Table \ref{tab:zerogmexps} summarizes the values of the atomic radiative lifetimes of the first
excited
$np$ states of Na, K, Rb and Cs which have been obtained through photoassociation of laser
cooled atoms in
the pure long-range $0_g^-$ state of the corresponding dimer, together with other recent
high-precision
measurements using various methods.

Table \ref{tab:zerogmsynth} displays the main characteristics of these different studies.
Since the
$0_g^-$ potential is a pure long-range potential, one does not need to find a way to deal with an
inner part of
the potential in order to avoid reducing the precision, as it would be the case for a "normal"
potential.
In all studies, the retardation effect has been introduced according to Equation
(\ref{eq:retard}). For
light atoms retardation has been found impossible to ignore: PAS of the $0_g^-$ state of
Na$_2$ provided
the first reported evidence of such effects in molecular spectra and an estimation of the
retardation
contribution for different alkali dimers \cite{jones1996}. In the case of heavier atoms, the
effect is
less important. For cesium, a fit without retardation yields parameters which are not
significantly
different from the ones of Ref \cite{bouloufa2007}, which included retardation. Besides
resonant dipole
and spin-orbit interaction, which are always the dominant terms, the terms of the multipole
development in
$R^{-6}$ and $R^{-8}$ are introduced in all studies, but the corresponding parameters, or
ratios between
them are sometimes kept fixed to a theoretical value (see Table \ref{tab:zerogmsynth}). The
$R^{-10}$ term
is generally negelected, except in Ref \cite{freeland2001}, where its influence is discussed.
The
influence of molecular rotation is accounted for in all studies, in the form given by Equation
(\ref{eq:Hrot}). Slightly different ways of dealing with the mixing of $\Pi$ and $\Sigma$
states have been
used, as we will see below. Asymptotic exchange interaction has to be introduced for the
heavier elements,
Rb and Cs. The variation with $R$ of the spin-orbit interaction was considered in Refs
\cite{gutterres2002,amiot2002}. Finally the validity of the Born-Oppenheimer approximation is
carefully
investigated in the sodium study \cite{jones1996}. Hyperfine structure does not appear in PA
spectra,
except for very high-lying levels. As it is neither resolved in the experiments nor introduced
in the
models used for the lifetime value extraction, it constitutes an important limitation to the
final
precision of the determinations.

\begin{table}[h]
\caption{\label{tab:zerogmexps}
\footnotesize Determinations from photoassociation spectroscopy (PA) of the radiative
lifetimes of the
atomic $np^2P_{3/2}$ states of Na, K, Rb and Cs (in bold face), together with recent
measurements by other
methods. We quote only references dating from 1994 or later; a good review on precision
lifetime
measurements on alkali atoms up to 1996 can be found in \cite{volz1996}. For the experiments
involving PA
of cold atoms, the atom cloud temperature is recalled, in parenthesis. For rubidium, the value
noted $^*$,
attributed to Ref \cite{boesten1997}, is calculated using their result for
$<s|r|p_{1/2}>*<s|r|p_{3/2}>$
together with the ratio of the $5p ^2P_{1/2}$ and $5p ^2P_{3/2}$ lifetimes of Ref
\cite{volz1996}. For
cesium, the value, noted $^{**}$, from Ref  \cite{derevianko2002}, is based on an experimental
determination of the $C_6$ van der Waals coefficient of the ground molecular state
\cite{chin2004b} and on
a theoretical relationship between this coefficient and the $C_3$ coefficients.}
  \begin{center}
  \begin{tabular}{c|*{4}{@{\hspace{.5cm}}c}@{\hspace{.5cm}}}
    \hline\hline
    \rule{0cm}{.2cm}
       element & author(date) & ref & $\tau_{3/2}$ (ns) & method    \\
    \hline
    \rule{0cm}{.2cm}
    Na($3s-3p$) & \bf{Jones {\it et al.} (1996)} 	 		& \cite{jones1996}
 	&
\bf{16.230(16)}	& {\bf PA} ($500~\mu$K)     \\
                & Oates {\it et al.} (1996) 				 	&
\cite{oates1996}      	
& 16.237(35)  		& linewidth       \\
                & Tiemann {\it et al.} (1996) 			 	& \cite{tiemann1996}
 	&
16.222(53)  		& mol spectr      \\
                & Volz {\it et al.} (1996) 				 		&
\cite{volz1996}
     	& 16.254(22)  		& fast beam       \\
    K($4s-4p$)  & \bf{Wang {\it et al.} (1997)} 		 	& \cite{wang1997b}  			
	& \bf{26.34(5)}   & {\bf PA} ($500~\mu$K)    \\
                & Volz {\it et al.} (1996) 				 		&
\cite{volz1996}
     	& 26.45(7)    		& fast beam       \\
    Rb($5s-5p$) & \bf{Freeland {\it et al.} (2001)} 	& \cite{freeland2001}   	&
\bf{26.24(7)} 	
& {\bf PA} ($700~\mu$K)     \\
                & \bf{Gutteres {\it et al.} (2002)} 	& \cite{gutterres2002}  	&
\bf{26.33(8)} 	
& {\bf PA} ($120~\mu$K)    \\
                & Volz {\it et al.} (1996) 				 		&
\cite{volz1996}      	
	& 26.24(4)    		& fast beam       \\
                & Boesten {\it et al.} (1997) 			 	& \cite{boesten1997}   	
	&
26.67(34)$^*$   & PA ($\sim$mK)             \\
                & Simsarian {\it et al.} (1998) 		 	& \cite{simsarian1998} 	
	&
26.20(9)    		& photon counting \\
    Cs($6s-6p$) & \bf{Amiot {\it et al.} (2002)} 	 		& \cite{amiot2002}
 	&
\bf{30.462(3)}  & {\bf PA} ($300~\mu$K)    \\
                & \bf{Bouloufa {\it et al.} (2007)} 	& \cite{bouloufa2007}   	&
\bf{30.41(30)}
& {\bf PA} ($300~\mu$K)  \\
                & Young {\it et al.} (1994) 				 	&
\cite{young1994}      	
& 30.41(10)   		& pulsed laser    \\
                & Rafac {\it et al.} (1999) 				 	&
\cite{rafac1999}       	
& 30.57(7)    		& fast beam       \\
                & Derevianko {\it et al.} (2002) 	 		&
\cite{derevianko2002}  	&
30.39(6)$^{**}$ & from C$_6$      \\
    \hline\hline
  \end{tabular}
\end{center}
\end{table}

\begin{table}[h]
\caption{\label{tab:zerogmsynth}
\footnotesize Comparison of the determinations of atomic lifetimes from photoassociative
spectroscopy of
the $0_g^-$ states. The most important points are recalled, when available. Displayed here
are: the
estimated experimental uncertainty on the measured energy of the vibrational states (in MHz),
$\sigma_E$;
the number of levels included in the fitting procedure, $N$; the smallest and largest values
of the
external Condon points of these levels, $R_{Cmin}$ and $R_{Cmax}$ (1$a_0=$0.0529277~nm); the
energy terms
that were included in the asymptotic potential in the considered analysis:
non-Born-Oppenheimer (non-BO),
exchange (exch.) or relativistic (rel.) terms;  the "effective" number of free parameters
included in the
fit, $p$ ({\it i.e.} the number of parameters of the model decreased by one each time either a
parameter
or a ratio between two parameters is kept fixed); the $\chi^2$ value of the fit (eventually
recalculated
to fit the definition of Equation (\ref{eq:chi2})), when it is available; the estimated
relative
uncertainty of the $np_{3/2}$ radiative lifetime, $\sigma_{\tau}$ (in $\%$), as given in the
original
reference. The values of $R_{Cmin}$ correspond to the bottom of the external well. The value
of $R_{Cmax}$
noted $^*$  for Na has been calculated from the value of the binding energy of the highest
level included
in the fit (similar data were not available for potassium); for Rb (resp. Cs), the values
of $R_{Cmax}$
are deduced from the corresponding $v$ value using \cite{fioretti2001} (resp.
\cite{fioretti1999}).}
  \begin{center}
  \begin{tabular}{c|*{10}{@{\hspace{.5cm}}c}@{\hspace{.5cm}}}
    \hline\hline
    \rule{0cm}{.2cm}
       atom & $\sigma_E$ (MHz) & $N$ & $R_{Cmin}$(a$_0$) & $R_{Cmax}$(a$_0$) & energy terms &
$p$ &
$\chi^2$ & $\sigma_{\tau}$($\%$) \\
    \hline
    \rule{0cm}{.2cm}
$^{23}$Na    	 \cite{jones1996}     & 5       & 7  & 70 & 122$^*$ & non-BO      & 2 &      &
0.1 \\
$^{39}$K  		 \cite{wang1997b}      & $<$60   & 23 & 52 &         & non-BO      & 2
& 0.42 &
0.2 \\
$^{85,87}$Rb   \cite{freeland2001}  & 60      & 75 & 32 & 90      & exch.       & 5 & 2.5  &
0.27 \\
$^{87}$Rb      \cite{gutterres2002} & 300     & 56 & 32 & 166     & exch., rel. & 8 & 1.1  &
0.3 \\
$^{133}$Cs     \cite{amiot2002}     & 150     & 75 & 23 & 65      & exch., rel. & 9 & 1.7  &
0.01 \\
$^{133}$Cs     \cite{bouloufa2007}  & 150     & 71 & 23 & 66      & exch., rel. & 6 & 0.41 &
1. \\
    \hline\hline
  \end{tabular}
\end{center}
\end{table}

In all studies, calculated vibrational energies are finally fitted to $N$ chosen experimental
data with
$p$ free parameters. Uncertainty is given for the lifetime value, which sometimes includes an
estimation
of systematic errors. Table \ref{tab:zerogmsynth} displays, when available, the experimental
uncertainty
$\sigma_E$, $N$, $p$ and the relative uncertainty on the lifetime value, $\sigma_{\tau}$. To
characterize
the $R$-range involved in the analysis, we present also, when it was possible to find them or
to calculate
them, the values of the external Condon points $R_{Cmin}$ and $R_{Cmax}$ of the lowest and
highest levels
included in the fit, respectively.

\subsection{Sodium}												

\label{sodium}

The $0_g^-$ external well of Na$_2$ is very shallow, with only 12 vibrational levels. The
energies of two
rotational levels ($J=2$ and $J=4$) for $v$ values ranging from 0 to 7 were measured with an
uncertainty
of 5~MHz. The analysis of the data was made by starting from the model of Movre and Pichler
\cite{movre1977} and by adding successively the most important corrections: retardation,
$R^{-6}$ and
$R^{-8}$ terms, non-adiabatic diagonal corrections and rotation. In addition to the
$B_vJ(J+1)$ term,
corresponding to the $\mathbf{J}^2$ term of Equation (\ref{eq:Hrot}), the authors calculated
the
$\mathbf{j}^2$ term by using the previously calculated mixing of $\Sigma$ and $\Pi$ states.

This progressive introduction of the energy terms allowed the authors to clearly point out the
role of the
different corrections, in particular of the retardation effect: they were able to give an
accurate
estimation of the contribution of this effect to the well depth. They also performed a
complete
coupled-channel calculation including all molecular potentials correlated to the ($3s+3p$)
limit, which
accounts for non Born-Oppenheimer effects. The long-range potentials were smoothly connected
at
$R=35$~a$_0$ to {\it ab initio} calculations. This complete calculation shows that the non
Born-Oppenheimer effects are very small. The influence of the inner part of the potentials was
found fully
negligible (using a hard wall at 35~a$_0$ gave almost no shift of the relevant eigenenergies).
A fit of
the $J=2$ results was made while keeping constant the two $C_8$ coefficients (using the values
of Ref
\cite{marinescu1995}) and the ratio between the two $C_6$ ones (using the same reference).
>From the result
of this fit, a value of the $3p$ atomic radiative lifetime was deduced, which is in good
agreement with
the result of fast beam measurements of Ref \cite{volz1996} and with two other recent
experimental values,
based on linewidth analysis and on molecular spectroscopy of low-lying vibrational levels (see
Table
\ref{tab:zerogmexps}). It also agrees well with the theoretical result of Ref
\cite{jonsson1996}. All
these data have completely removed a long-standing discrepancy between experiment and theory
(see
\cite{volz1996}).

\subsection{Potassium}												

\label{potassium}

A PAS study with ultracold potassium atoms was performed in 1997 \cite{wang1997b}, with a "dark
spot"
magneto-optical trap. The vibrational levels of the pure long-range $0_g^-$ potential
converging to the
($4s+4p_{3/2}$) limit have been observed between $v=0$ and $v=30$. The rotational component
$J=2$ of most
levels was measured with an uncertainty smaller than 60~MHz. The analysis is performed along
the same
lines as for sodium. The calculated rovibrational energies are fitted to 23 measured ones
($v=0-12,14-17,19-22,24,26$); the ratios between $C_6$ and $C_8$ coefficients are kept fixed
to a
theoretical value \cite{marinescu1995}. The one-$\sigma$ standard deviation is 0.0013
cm$^{-1}$.
The uncertainty on the $C_{3}$ value includes both a part coming from the fitting procedure
and a
systematic part coming from the limitations of the model, essentially from neglecting the
molecular
hyperfine structure.
The estimation of the final uncertainty is however not described in detail. The lifetime value
agrees well
with the measurement of Ref \cite{volz1996} (see Table \ref{tab:zerogmexps}).

\subsection{Rubidium}												

\label{rubidium}

There are two independant studies \cite{freeland2001,gutterres2002}, using different
experimental data.

The first experiment \cite{freeland2001} used a FORT trap with a temperature of about
700~$\mu$K and
doubly spin-polarized atoms. From photassociation spectra of both $^{85}$Rb and $^{87}$Rb
atoms, a large
number of rovibrational energy levels of the $0_g^-$ potential of both isotopes ($v=0-40$ for
$^{85}$Rb
and $v=0-24,28-35,41-47$ for $^{87}$Rb) were obtained with an uncertainty of the order of
60~MHz. The
diagonalization of the matrix of Equation (\ref{eq:MPplus}) was done analytically on a
simplified form of
the matrix, before the addition of the correcting terms, some of them (non-adiabatic terms,
rotation)
depending on the mixing of states which is characterized by the eigenvectors of the
diagonalization. The
terms introduced are: retardation effect on the resonant dipole interaction, dispersion terms
up to
$R^{-8}$, asymptotic exchange interaction, rotation (using the known mixing of $j=0$ and $j=2$
states).
Finally, the effect of $R^{-10}$ terms  and of non-adiabatic corrections have been tested. A
rather
detailed study of these effects and of their influence on the results of the fit procedure is
given: we
will comment on this point in the last section.

The other study \cite{gutterres2002} has been performed on the experimental data of Ref
\cite{gabbanini2000}, concerning only $^{87}$Rb, in a MOT trap at about 120~$\mu$K. The
analysis of the
data is conducted in a similar way. The $v$ values included in the fit were
$v=0,1,12-23,30-71$. The
diagonalization is done numerically on the complete form of the two-state matrix given by
Equation
(\ref{eq:MPplus}), including retardation effect on the resonant dipole interaction, dispersion
terms up to
$R^{-8}$, asymptotic exchange interaction, rotation, intra-atomic relativistic corrections and
$R$-varying
spin-orbit terms. Concerning rotation, the $\mathbf{j}^2$ term was simply calculated using the
asymptotic
value $j=2$. The spin-orbit $R$-variation, which was suggested by quantum chemistry
calculations, appeared
to greatly diminish the agreement between experimental and calculated values. The estimation
of the
uncertainty on the lifetime value will also be discussed in the last section.

The lifetime values obtained from the two PAS studies are fully compatible, and in agreement
with most of
the previous experimental determinations (see Table \ref{tab:zerogmexps}).

\subsection{Cesium}												

\label{cesium}

In the cesium case, the two published studies \cite{amiot2002,bouloufa2007} from our group
concern the
analysis of the same PAS data from ref.\cite{fioretti1999}, from which a Rydberg-Klein-Rees
(RKR)
potential was previously extracted. Our second analysis \cite{bouloufa2007} was necessary in
order to
solve remaining discrepancies in the intensity of the PA spectrum and in the scattering length
and the van
der Waals ground state $C_6$ values that we deduced in Ref \cite{drag2000}.

The $0_g^-$ potential of Cs$_2$ is not strictly speaking a pure long-range one: the minimum of
the
external well is located around 25~a$_0$, whereas the LeRoy criterion which is generally used
for the
definition of long-range distances \cite{ji1995} yields a distance of about 28.5~a$_0$. It was
therefore
unavoidable to introduce the asymptotic exchange term. The top of the potential barrier
separating the
internal and external well is critically related to the exchange term and is expected to be
close to the
energy of the dissociation limit \cite{vatasescu2000}. The imperfect knowledge of the height
of the
barrier will affect the description of the highest levels, which were therefore not introduced
in the fit.
Relativistic atomic corrections were introduced and, in the first paper, $R$-variation of the
spin-orbit
was also considered, like in Ref \cite{gutterres2002}. Both studies used exactly the same
model and the
same least-square fitting code.

The main difference between the two cesium studies was the $v$-labeling of the observed
levels: the level
numbered $v=0$ in the first reference \cite{amiot2002} was labeled $v=2$ in the second one
\cite{bouloufa2007}.
The change in the $v$ labeling affected the shape of the bottom of the $0_g^-$ potential (it
is of course
deeper, to admit two more levels), but its long-range part remained unchanged. As a
consequence, the two
lifetime values are close to each other and both compatible with the measurement of Ref
\cite{young1994}
rather than with the one of Ref \cite{rafac1999} (see Table \ref{tab:zerogmexps}). However, we
took the
opportunity of the second study to investigate more carefully the estimation of the error
bars; taking
then into account the correlations between the parameters, we found an uncertainty for the
lifetime value
strikingly larger than in the previous study. We will come back to this point in the last
section.

\section{Determination of radiative lifetimes of L\lowercase{i}, C\lowercase{a},
S\lowercase{r},
Y\lowercase{b}
atoms from long-range analysis}
\label{sec:longrange}

As shown in Section \ref{sec:LRB}, atomic interaction parameters can be obtained from a
careful analysis
of the energy of the high-lying molecular levels. McAlexander {\it et al.}
\cite{mcalexander1995,mcalexander1996} reported two studies of the $A^1\Sigma_u^+$ potential
of Li$_2$
converging to the ($2s+2p$) limit while the $B^1\Sigma _u^+$ potentials converging to the
$((ns^2)^1S_0+(nsnp)^1P_1)$ limit have been investigated for Ca \cite{vogt2007}, Sr
\cite{nagel2005,yasuda2006} and Yb \cite{takasu2004} (with $n$=3, 4 and 6 respectively).

Table \ref{tab:longrangeexps} shows the results obtained from PAS by different groups,
together with
recent atomic lifetime measurements obtained by other methods. Table \ref{tab:longrangesynth}
displays the
main characteristics of the PAS studies.
The last two rows refer to LeRoy-Bernstein fits, in which only the $R^{-3}$ term is included.
It is emphasized here that the straightforward approach of directly fitting the data to
the LeRoy-Bernstein
formula, although quite tempting for those new to the method, can lead to misleading data,
even when the
fit seems quite good. Modified forms of the LeRoy-Bernstein law, involving a larger number of
parameters,
are proposed in Ref \cite{comparat2004,jelassi2008}. In the latter, the validity criteria are
given in
terms of the energy value, and not in terms of interatomic distance, as it is usually done.
For the other cases, we indicate in the Table how the authors managed the inner part of the
potential.
Rotation and retardation are always introduced in the manner described in Section
\ref{sec:dipole-dipole}. Dispersion terms in $R^{-n}$ with $n=6$ are introduced in the Ca
study, and with
$n$=6, 8 in the Li study. Non Born-Oppenheimer effects are also included in the latter.

\begin{table}[h]
\caption{\label{tab:longrangeexps}
\footnotesize Determinations from photoassociation spectroscopy (PA) of the radiative
lifetimes of the
atomic $2p~^2P_{1/2}$ state of Li and of the $(nsnp)~^1P_{1}$ states of Ca, Sr and Yb (bold
characters),
together with recent measurements by other methods (normal characters). The two molecular
asymptotes
involved in the experiment are recalled in the first column. Reference \cite{blagoev1994}
contains 6
experimental values, obtained with various methods.}
  \begin{center}
  \begin{tabular}{c|*{5}{@{\hspace{.5cm}}c}@{\hspace{.5cm}}}
    \hline\hline
    \rule{0cm}{.2cm}
       element & author(date) & ref & $\tau$ (ns) & method    \\
    \hline
    \rule{0cm}{.2cm}
        Li                   		    				&\bf{McAlexander {\it
et al.}
(1996)}&\cite{mcalexander1996}&\bf{27.102(7)} 		&{\bf PA} ($<$1~mK)    \\
\footnotesize{$(2s)^2S_{1/2}-(2p)^2P_{1/2}$}&Linton {\it et al.} (1996)
&\cite{linton1996}
&27.09(8)       		&FT mol.spectr.\\
                                            &Volz {\it et al.} (1996)
&\cite{volz1996}
&27.11(6)       		&fast beam     \\
        Ca                                  &\bf{Vogt {\it et al.} (2007)}
&\cite{vogt2007}
&\bf{4.639(2)}$^*$  &{\bf PA} (1.5~mK)      \\
\footnotesize{$(4s^2)^1S_0-(4s4p)^1P_1$}    &Hansen {\it et al.} (1983)
&\cite{hansen1983}
&4.60(20)       		&photon counting     \\
					    									
						&Kelly {\it et al.} (1980)
&\cite{kelly1980}
 &4.49(7)        		&Hanle effect  \\
        Sr  				  									
		&\bf{Nagel {\it et al.} (2005)}      &\cite{nagel2005}      &\bf{5.22(3)}   		
&{\bf PA} (2~mK)       \\
\footnotesize{$(5s^2)^1S_0-(5s5p)^1P_1$}    &\bf{Yasuda (2006)}   						
	 &\cite{yasuda2006}     &\bf{5.263(4)}  		&{\bf PA} ($\sim\mu$K) \\
              				    									
&Lurio {\it et al.} (1964)           &\cite{lurio1964}      &4.97(15)       		
&Hanle effect  \\
              				    									
&Kelly {\it et al.} (1980)           &\cite{kelly1980}      &4.68(10)       		
&Hanle effect  \\
        Yb 				    									
		&\bf{Takasu {\it et al.} (2004)}   	 &\cite{takasu2004}   	&\bf{5.464(5)}  		
 &{\bf PA} (40~$\mu$K)  \\
\footnotesize{$(6s^2)^1S_0-(6s6p)^1P_1$}    &Blagoev {\it et al.} (1994)
&\cite{blagoev1994}
&compilation        & several\\
    \hline\hline
  \end{tabular}
  \end{center}
\end{table}

\begin{table}[h]
\caption{\label{tab:longrangesynth}
\footnotesize Similar to Table II, but for potentials that are not "pure long-range". There is
thus a new
column qualifying the inner part of the potential. The inclusion of $R^n$ terms, of rotation
(rot.),
retardation (ret.), and non-Born-Oppenheimer (non-BO) terms is specified. The names of the
different
parameters involved in the fitting procedure are specified (with "node pos." standing for
nodal line
position and "int.wall" for repulsive wall position). The values of $R_{Cmin}$ (resp.
$R_{Cmax}$) noted
$^*$ have been calculated from the values of the binding energy of the lowest (resp. highest)
level
included in the fit. The two isotopes of lithium are studied in the same reference; the
authors did two
separate fits, then a weighted average. The experimental uncertainty $\sigma_E$ was not
reported in Refs
\cite{mcalexander1996} and \cite{takasu2004}.}
  \begin{center}
  \begin{tabular}{c|*{10}{@{\hspace{.45cm}}c}@{\hspace{.45cm}}}
    \hline\hline
    \rule{0cm}{.2cm}
       atom & $\sigma_E$ & $N$ & $R_{Cmin}$ & $R_{Cmax}$&inner part&asymptotic part&
parameters & $\chi^2$
& $\sigma_{\tau}$ \\
  	     & {\small (MHz)}&    & {\small (a$_0$)}& {\small (a$_0$)}& &            &
   &
  & {\small ($\%$)}          \\
    \hline
    \rule{0cm}{.2cm}
$^6$Li \cite{mcalexander1996} &     & 23  & 29     & 150     & RKR           &
rot.,ret.,non-BO,  &
$D,C_3,C_6$   &      & 0.026 \\
$^7$Li \cite{mcalexander1996} &     & 27  & 30     & 170     &               &
$R^{-6}$,$R^{-8}$  &
         &      & \\
$^{40}$Ca \cite{vogt2007}     & 10  & 8   & 83     & 127     & nodal line    &
rot.,ret.,$R^{-6}$ &
$C_3$,node pos. &      & 0.037 \\
$^{88}$Sr \cite{nagel2005}    & 5   & 14  & 380    & 605     &{\it ab initio}& rot.,ret.
   &
$C_3$,int. wall & 0.79 & 0.57 \\
$^{88}$Sr \cite{yasuda2006}   & 300 & 62  & 60     & 208     &               & LR-B
   &
$C_3,v_D$       &      & 0.076 \\
$^{174}$Yb \cite{takasu2004}  &     & 72  & 60$^*$ & 185$^*$ &               & LR-B
   &
$C_3,v_D$       &      & 0.09 \\
    \hline\hline
  \end{tabular}
  \end{center}
\end{table}

\subsection{Lithium}												

\label{lithium}

For lithium, the spin-orbit interaction between the attractive $\Sigma$ state and the
repulsive $\Pi$
state is much weaker than for the other alkalis and cannot compete with the resonant dipole
interaction to
give rise to a pure long-range potential well. In order to calculate vibrational energies and
wavefunctions, the asymptotic part of the potential has to be completed by a description of
its inner
part. In the study of Ref \cite{mcalexander1996}, the A$^1\Sigma_u^+$ potential curve was
constructed
using the RKR potential of Ref \cite{linton1996} for the inner part, and, for the outer part,
using an analytic
form of the long-range interaction. The depth of the potential $D$ is considered as an
adjustable
parameter. The RKR potential is extrapolated at short distance with two {\it ab initio} points
and is
smoothly connected, at about 25.4~a$_0$, to the long-range interaction, which includes the
$C_3/R^3$,
$C_6/R^6$ and $C_8/R^8$ terms of the multipole expansion and the first-order corrections to
the
Born-Oppenheimer approximation. The retardation effects are introduced as in the previous
section. The
hyperfine structure of the lines was calculated by first-order perturbation theory in order to
precisely
locate the center of gravity within each observed vibrational level. The vibrational energies
introduced
in the fit correspond to $v=63-72,76-88$ for $^6$Li and $v=69-76,79-97$ for $^7$Li. Separate
fits were
performed for the two isotopes, followed by a weighted average. The lifetime value is
remarkably accurate
(0.026 $\%$); it is in good agreement with the other, less precise, experimental values of
Table
\ref{tab:longrangeexps}, and in excellent agreement with the very accurate {\it ab initio}
calculated
values (see \cite{yan1995} and several other values quoted in \cite{mcalexander1996}). As
claimed by
McAlexander {\it et al.}, the precision of their analysis was sensitive to non
Born-Oppenheimer effects,
to retardation effects and to relativistic effects in atomic structure calculations.

\subsection{Calcium}												

\label{calcium}

The experiment was performed with calcium atoms in a MOT at a temperature of about 1.5~mK
\cite{vogt2007}.
The PAS lines corresponding to 8 rovibrational levels  ($J=1,3$ and $v^*=69,72,80,85$) of the
excited B
$^1\Sigma_u^+$ potential converging to the $((4s^2)^1S_0+(4s4p)^1P_1)$ limit were observed and
analyzed,
with a final experimental uncertainty for the level energies $\sigma_E \sim 10$~MHz. In order
to account
for line shifts and broadening induced either by the finite temperature of the atom cloud or
by the power
of the PA laser, the atom trap loss was carefully modelled. The authors used the formalism of
Bohn and
Julienne \cite{bohn1999}, which yields the temperature dependence of the PA profile once both
ground and
excited potential are known. The ground state potential was taken from Ref \cite{allard2003}.
For the
excited potential, they used an asymptotic part including resonant dipole interaction
with
retardation correction, $R^{-6}$ and rotation terms
\begin{eqnarray}
V(R)=D-\frac{C_3}{R^3}f^{\Sigma}(R)-\frac{C_6}{R^6}+\frac{\hbar^2[J(J+1)+2]}{2\mu R^2},
\label{eq:potCa}
\end{eqnarray}
where the rotation term is obtained from Equation (\ref{eq:Hrot}), where the $j=1$ value of
the electronic
angular momentum is taken into account. Instead of using a defined potential in the inner
part, the
authors fixed boundary conditions near the frontier of the long-range region ($R=0.95$ nm). They
imposed on
the vibrational wavefunctions to vanish on a nodal line whose position was taken as an
adjustable
parameter, according to the accumulated phase method of Refs
\cite{crubellier1999,vanhaecke2004}. This
method was checked to give the same results as the one in which the asymptotic part is
connected to an
{\it ab initio} potential \cite{allard2004} with an adjustable repulsive wall. An iterative
procedure was
used, since the parameters of the asymptotic potential required to analyze the profiles were
deduced from
the result of this same analysis. The very precise (0.04 $\%$) lifetime value that they
obtained for the
atomic level $^1$P$_1$ is found in agreement with the value obtained by photon counting
\cite{hansen1983},
but not with the one based on the Hanle effect \cite{kelly1980} (see Table
\ref{tab:longrangeexps}). It
agrees well with the many-body calculations of Ref \cite{porsev2001}, but not so well with the
quantum
chemistry ones \cite{bussery-honvault2006} or with the Multi-Configuration Hartree-Fock ones
\cite{fischer2003}.

\subsection{Strontium}												

\label{strontium}

Two different groups reported measurements of PAS in cold strontium
\cite{nagel2005,yasuda2006}.
Both were using $^{88}$Sr in a magneto-optical trap, but with rather different temperatures
(see Table
\ref{tab:longrangeexps}). In both experiments only $s$-wave collisions are expected to occur
so that only
a single rotational level $J=1$ is excited.

In the first experiment \cite{nagel2005}, the energy of the vibrational levels with $v^*$
ranging from 48
to 61 was claimed to be measured with an uncertainty $\sigma_E$ of the order of 5~MHz. The
analysis was
made by quantum calculations.
The asymptotic behavior of the potential was given by Equation (\ref{eq:potCa}) (without the
$R^{-6}$
term) and was smoothly connected, at $R=1.5$~nm, to an {\it ab initio} potential
\cite{boutassetta1996}.
The position of the inner wall was considered as an adjustable parameter. Their best fit was
characterized
by $\chi^2=0.79$.

The second experiment \cite{yasuda2006} had a larger experimental uncertainty
$\sigma_E\sim$~300~MHz. The
measurements concerned levels with $v^*=84-134,150-160$ and the analysis was made using
LeRoy-Bernstein
law.

The two results are not strictly speaking compatible (the gap between the two confidence
intervals is
larger than the error bar of the second reference). Neither result agrees with previous
measurements (see
Table \ref{tab:longrangeexps}); the agreement is better with calculations of Ref
\cite{porsev2001}.

\subsection{Ytterbium}												

\label{ytterbium}

Ytterbium is a rare-earth element with electronic structure in the ground state
$4f^{14}6s^{2}$ similar to
the one of alkaline-eath atoms.
The atoms were prepared in a FORT trap at a temperature of about 100~$\mu$K \cite{takasu2004}.
About 72
levels ($J=1$ and $v^*=103-174$) were measured  and assigned to the $^1\Sigma _u^+$ potential
converging
to the $(^1S_0+^1P_1)$ limit. The analysis was made using the LeRoy-Bernstein law. Rotation is
expected to be
extremely weak and was not introduced. The residuals of the fit are less than 0.5 $\%$. The
influence of
the neglected effects is estimated and the main limitation of the precision is claimed to be
the line
broadening due to predissociation. The extremely precise value which is obtained is in
agreement with most
of the much less accurate previous measurements.

\section{Discussion of accuracy issues}
\label{accuracy}

The key advantage of determining atomic lifetimes from photoassociative spectroscopy is that
their values
are deduced from high resolution molecular spectroscopy. However, transmitting this precision
to the
atomic radiative lifetime value is not a trivial matter. In the following, we describe in
detail how the
quality or the "goodness" of the fit and the confidence interval of the optimized parameters
should be
properly investigated. We illustrate our derivation through a numerical application within a
linear
approximation applied to some of the experiments reported in the previous section.

\subsection{Accuracy of a parameter determination from a fit procedure}
\label{sec:precisiongene}

The quality of the fit is primarily characterized by the minimum value of the least-square
$\chi ^2$
function
\begin{eqnarray}
\chi^2=\frac{\sum_{i=1,N}(E_{calc}-E_{exp})^2}{(N-p)\sigma_E^2},
\label{eq:chi2}
\end{eqnarray}
which should be close to one, or by the rms value,
\begin{eqnarray}
rms=\sqrt{\frac{\sum_{i=1,N}(E_{calc}-E_{exp})^2}{(N-p)}},
\label{eq:rms}
\end{eqnarray}
which has the dimension of an energy and has to be close to $\sigma_E$. It is worth mentioning
that the
estimate of the uncertainty on the measurements is most generally not well known: one often
uses the
results of the fit to define an "unbiased" value of this error. One can get further
information on the
quality of the fit by analyzing the residuals (see for instance \cite{femenias2003}). When the
data have a
natural order, like it is the case here, the non-stochastic trend of their distribution can be
checked
visually, or by a more elaborate method. We tried the method of Ref \cite{femenias2003} as an
a posteriori
test for the residuals of our cesium study \cite{bouloufa2007}: the frequency of sign changes
was found
to be 0.3286, whereas the ideal value corresponding to $p=6$ free parameters was 0.5269 with a
variance of
0.05917. According to this criterion, our fit was therefore not completely satisfying. This
was indeed
qualitatively visible in a graph of the ordered distribution of the residuals (not shown in
our paper).

Once the fit has been checked to be unbiased, one has to evaluate the error bars on the
parameter values.
In all cases of interest here, the least-mean square function $\chi^2$ is a complicated
non-linear
function of the different parameters. However, close to the best fit region, it is often
possible to
linearize the model, {\it i.e.} to consider that the calculated energies $E_v$ depend
approximately
linearly on the parameters (see Ref \cite{vanhaecke2003b} and references therein, in
particular Ref
\cite{group2000}). Let us call $\mathbf{X}$ the $N \times p$ matrix of the derivatives of  the
$N$
calculated $E_v$ values with respect to the $p$ parameters $a_i$, with
\begin{eqnarray}
\mathbf{X}_{v,i}=\frac{\partial E_v}{\partial a_i}.
\label{eq:matderivees}
\end{eqnarray}
The theory of linear regression can then be used, with the $\mathbf{X}$ matrix playing the
role of the
model matrix $\mathbf{M}$ which relates the calculated energies to the parameters through the
vector
equation $\mathbf{E}=\mathbf{M}\mathbf{a}$,
where $\mathbf{E}$ is the $N$-dimensional vector of the $E_v$ values and $\mathbf{a}$ the
$p$-dimensional vector
of the parameters.
In particular the square of the one-parameter standard errors are the diagonal matrix elements
of the
$p\times p$ covariance matrix $\mathbf{V}$,
\begin{eqnarray}
\mathbf{V}=\sigma_E^2(\mathbf{X}^T~\mathbf{X})^{-1}.
\label{eq:cov}
\end{eqnarray}
where $\mathbf{X}^T$ is the transpose of the matrix $\mathbf{X}$.
The great interest of such a treatment is that it accounts for the correlations between the
parameters.

It is also possible to consider the case where only a part of the parameters of the model are
optimized
whereas some others are fixed to a value with a known uncertainty (see the PhD thesis of
Nicolas Vanhaecke
\cite{vanhaecke2003b}). Let be $\mathbf{a^*}$ (resp. $\mathbf{\tilde a^*}$) the vector of the
optimized
(resp. non-optimized) parameters at the minimum of $\chi^2$. The values of the optimized
parameters are
expected to change if the values of the non-optimized ones are taken at a value
$\mathbf{\tilde a}$
different from $\mathbf{\tilde a^*}$. Within the linear approximation, the value of the
optimized
parameters can be calculated without performing a new fit, according to
\begin{eqnarray}
\mathbf{a}= \mathbf{a^*}+ \big( \mathbf{X}^T \mathbf{X} \big) ^{-1} \mathbf{X} ^T \Big[
\mathbf{e} -
\mathbf{\tilde X} (\mathbf{\tilde a}-\mathbf{\tilde a^*})\Big],
\label{eq:modpar}
\end{eqnarray}
where $\mathbf{e}$ is the vector of the residuals.
It is possible to evaluate the error made on a given (adjusted) parameter value due to the
uncertainty of
the other (fixed) parameter.
We call $\mathbf{X}$ (resp. $\mathbf{\tilde X}$) the matrices of derivatives for the optimized
(resp.
non-optimized) parameters taken separately, and $\mathbf{V}$ the covariance matrix of the
optimized
parameters, according to Equation (\ref{eq:cov}). By analogy, we call $\mathbf{\tilde V}$ the
matrix whose
diagonal elements are the square of the uncertainties of the non-optimized parameters. The
restriction to
the optimized parameters of the total covarance matrix can be written as
\begin{eqnarray}
\mathbf{V_t}= \mathbf{V}+ \big( \mathbf{X}^T \mathbf{X} \big) ^{-1} \mathbf{X} ^T
\mathbf{\tilde X}
\mathbf{\tilde V} \mathbf{\tilde X} ^T \mathbf{X} \big( \mathbf{X}^T \mathbf{X} \big)^{-1}.
\label{eq:properr}
\end{eqnarray}
If the model is not close enough to a linear one, the most direct way to account for the
correlations
between the parameters is to draw $\chi^2$ contours, corresponding to the minimum $\chi^2$
values obtained
by varying step by step one particular parameter while letting all other free.  Different
conditions,
based for instance either on the Fisher distribution with $p$ and $N-p$ degrees of freedom or
on the so
called $\chi^2$ law (also called Pearson law with $N-p$ degrees of freedom) allow one to find
conditions
for the $\chi^2$ values defining the confidence ellipsoid corresponding to the chosen parameter
(see for
instance Ref \cite{group2000}).

\subsection{The LeRoy-Bernstein law}										
\label{precisionLRB}

We first consider the case of the strontium study \cite{yasuda2006} and of the ytterbium study
\cite{takasu2004}, where the data are fitted to the LeRoy-Bernstein (LR-B) law with two
parameters, $C_3$
and $v_D$. We will assume here that the model can be linearized.

The goodness of the fit can be checked by analysing the residuals (see for instance Ref
\cite{femenias2003}). In the LR-B study of strontium \cite{yasuda2006}, a qualitative check of
the
residuals is possible and seems to be satisfying, if one assumes that the experimental uncertainty
$\sigma_E$ is
constant. It is more difficult to conclude this in the case of the LR-B study of Ytterbium
\cite{takasu2004};
the residuals are not shown and the authors claim that the deviations are everywhere smaller
than 0.5\%.
This might however imply much larger deviations for low lying levels, which are still probably
measured
with the same or higher precision (in absolute value): this could be the signature of a deviation
from LR-B
law for these levels.

Writing the LeRoy-Bernstein law in the simple form of Equation (\ref{eq:lrbter}),
one finds that the matrix elements of $\mathbf{H}=\mathbf{X}^T~\mathbf{X}$ are
\begin{eqnarray}
H_{11} & = & \sum_{v^*=v^*_a}^{v^*_b} (v^*-v^*_D)^{12} \\
H_{12}=H_{21} & = & -6X_0 \sum_{v^*=v^*_a}^{v^*_b} (v^*-v^*_D)^{11} \\
H_{22} & = & 36(X_0)^2\sum_{v^*=v^*_a}^{v^*_b} (v^*-v^*_D)^{10},
\label{eq:hesselmts}
\end{eqnarray}
where $v^*_a$ and $v^*_b$ are the limits of the $v^*$-values introduced in the fit (it is
assumed in the
above formulas that the $v^*$ values are contiguous, but it is straightforward to extend them
to any set
of $v^*$ values). Assuming that the least mean square function is locally linear
\cite{vanhaecke2003b},
the standard uncertainties on the two parameters are obtained from the diagonal matrix
elements of the
inverse of $\mathbf{H}$. Assuming now that the uncertainty of the energy level measurements
$\sigma_E$ is
constant, the standard error of $X_0$ is found to be
\begin{eqnarray}
\delta(X_0)^2=f(v^*_a,v^*_b) \sigma_E^2,
\label{eq:standarderror}
\end{eqnarray}
where $f(v^*_a,v^*_b)$ is a function of the extreme values of $v^*$ only, given by
\begin{eqnarray}
f(v_a,v_b)=\frac{H_{22}}{H_{11}H_{22}-H_{12}^2}.
\label{eq:fv}
\end{eqnarray}
The relative uncertainty of the lifetime value is thus
\begin{eqnarray}
\sigma_{\tau}=\frac{\delta(\tau)}{\tau}=\frac{1}{2}\frac{\delta(X_0)}{X_0};
\label{eq:relstandarderror}
\end{eqnarray}
It is of course proportional to $\sigma_E$ and, apart from its dependence on the extreme
$v^*$-values (see
Equation (\ref{eq:fv})), it is proportional to $(C_3)^2$ and to $\mu^3$ (see Equation
(\ref{eq:x0})).

A numerical application of the formula (\ref{eq:relstandarderror}) can be performed with the
characteristics
of the strontium study \cite{yasuda2006}, with $v^*$ values in the interval $v^*_a\sim84$ to
$v^*_b\sim160$ and an experimental uncertainty $\sigma_E \sim$ 300~MHz. The relative
uncertainty on the
lifetime value depends very little on the $v^*_D$ value, which is not given in the reference.
In the example
below, it varies by about 4 \% of its own value for $v^*_D$ varying from 0 to 1. However, as
we will see
below, it depends strongly on the extreme $v^*$-values. We find here an uncertainty
$\sigma_{\tau}$ of the
order of 0.088 \%, {\it i.e.} about the same as given in the paper.
For ytterbium \cite{takasu2004}, the extreme $v^*$ values are $v^*_a=103$ and $v^*_b=174$ but
the
experimental uncertainty is not given. The error bar given in the paper, 0.09 $\%$, would
correspond to
$\sigma_E \sim 200$~MHz, which is likely for such experiments.

\begin{figure}[t]
\includegraphics[width=30pc]{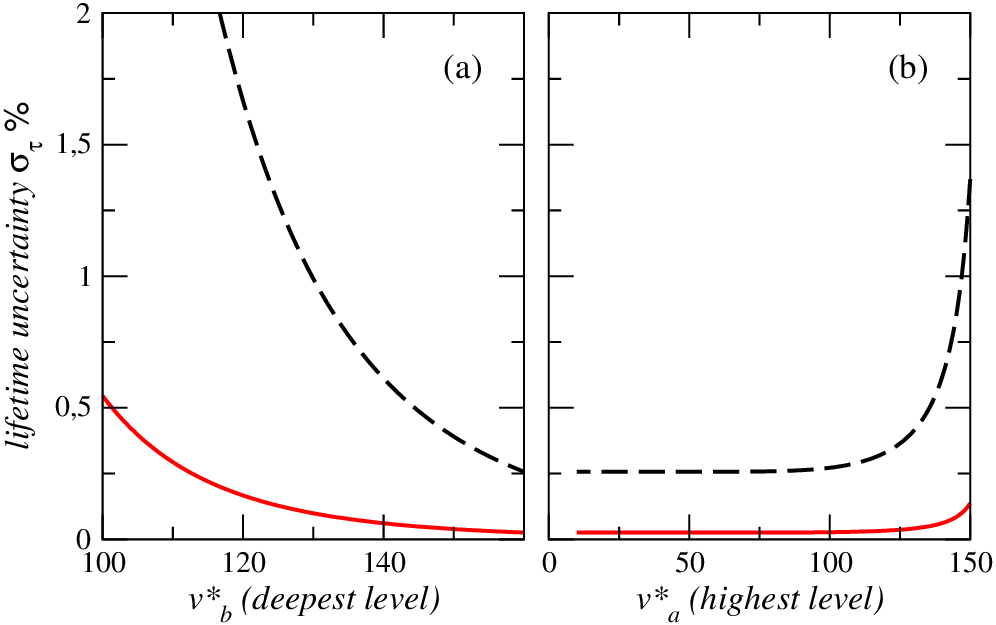}
\caption{\footnotesize Variation of the relative uncertainty $\sigma_{\tau}$ on the lifetime
value derived
from an analysis using LeRoy-Bernstein law. The chosen example is the one of $^{88}$Sr studied
in Refs
\cite{nagel2005} and \cite{yasuda2006}. The experimental uncertainty on the level energy
measurements is
assumed to be $\sigma_E=300$~MHz (dashed lines) or $\sigma_E=30$~MHz (full lines). In the left
graph, we
show the variations of $\sigma_{\tau}$ vs the $v^*$ value of the deepest level included in the
fit,
$v^*_b$; the $v^*$ value of the highest level is kept constant at $v^*_a=10$. In the right
graph, the
roles of $v^*_a$ and $v^*_b$ are exchanged; the fixed value is $v^*_b=160$. It is of course
assumed that
the LeRoy-Bernstein law is valid down to this $v^*_b$ value.}
\label{fig:erreurs}
\end{figure}

A general trend of the error bar on the lifetime value obtained from a two-parameter
LeRoy-Bernstein fit
can be illustrated on the strontium example. In Figure \ref{fig:erreurs}, we show the values
of
$\sigma_{\tau}$ corresponding to either a fixed $v^*_a$ value (lowest level) or a fixed
$v^*_b$ value
(highest level), the other $v^*$-limit varying. When fixing $v^*_a$ at 160, a very small
uncertainty of
the order of 0.25 $\%$ is obtained as soon as $v^*_b$ is of the order of 100. Conversely, even
for a value
of $v^*_a$ as low as 10, $\sigma_{\tau}$ does not approach this value before $v^*_b$ is very
close to 160.
One would say that the deepest levels are crucial to reduce the uncertainty for the $X_0$
parameter of the
LR-B fit. It is important to recall that it is assumed that the LR-B law is verified for all
the
considered $v^*$ values, which settles of course a lower bound to $v^*_a$. These results help
us to
qualitatively understand why the very precise measurements from $v^*=48$ to $v^*=61$ with
$\sigma_E=5$~MHz
of the first strontium study \cite{nagel2005} yielded a less accurate lifetime value than the
measurements
of the second one \cite{yasuda2006}, whose experimental uncertainty, $\sigma_E=300$~MHz, is
much larger,
but for wich the $v^*$ values run between $v^*=84$ to $v^*=160$.

It is interesting to notice that the present estimation of the error bar using the LeRoy-Bernstein
law is
meaningful for all studies of Section \ref{sec:longrange} ({\it i.e.} all studies but the
$0_g^-$ ones),
even if the analysis relies on quantum calculations. For calcium, using the $v^*$ values and
the
$\sigma_E$ value of Ref \cite{vogt2007} we find $\sigma_{\tau}\sim 0.033~\%$ instead of
$0.037~\%$; for
strontium, using the $v^*$ values and the $\sigma_E$ value of Ref \cite{nagel2005}, we find
$\sigma_{\tau}\sim 1.07~\%$ instead of $0.57~\%$. It is not surprising that the LR-B
estimation gives a
good result: as only two parameters are included in the fit, the situation is the same as
in a
LeRoy-Bernstein analysis. In the lithium studies, using the $v^*$ values of Ref
\cite{mcalexander1996} and
assuming $\sigma_E \sim 100$~MHz gives $\sigma_{\tau}\sim 0.025~\%$ for $^6$Li and
$\sigma_{\tau}\sim
0.11~\%$ for $^7$Li, instead of $\sigma_{\tau}\sim 0.026~\%$ for the two isotopes in the above
reference.
We did not expect such a good agreement, since three parameters are introduced in the model;
it appears
that the correlations between the $C_3$ and $C_6$ parameters (which are not included in our
estimation) do
not increase the uncertainty on  the $C_3$ value.

\subsection{Other cases}											

\label{precisionautres}

When the number of parameters increases, the matrix of the derivatives generally does not have
a
simple
analytical expression. In the different studies of PAS on alkalis reviewed here, it is often
difficult to
retrieve how the different authors evaluate the given final error bar. The experimental
uncertainty
$\sigma_E$ is sometimes missing as well as the $\chi^2$ or $rms$ value characterizing their
best fit and
the corresponding residuals.

In the sodium case, we did not find the $\chi^2$ or $rms$ value of the fit (it could be
recalculated,
since calculated and measured values of the 7 vibrational levels are displayed), and neither the
method used
for the estimation of the uncertainty. Systematic errors coming from non-optimized parameters
or from
insufficiencies of the model were examined.

In the potassium case, the situation is similar. The $\chi^2$ value given in Table
\ref{tab:zerogmsynth},
0.42, is derived from the $rms$ value of fit, 1.3x10$^{-3}$~cm$^{-1}$. Systematic errors are
introduced,
but it is difficult to find out wether the correlations between the 2 parameters are taken
into account or
not.

Concerning the first rubidium study \cite{freeland2001}, the $\chi^2$ value given in the
Table, 2.5, has
been recalculated to fit the definition of Equation (\ref{eq:chi2}) and might be considered as
being a
little too large. The estimation of the error bar on the final $C_3$ value is well described.
Systematic
errors are checked and correlations between parameters are in principle accounted for since
the author
draws the contours of the $\chi^2$ function just as described above (end of Section
\ref{sec:precisiongene}). However, we remark that some free parameters did not move (or moved
extremely
little) from their initial value. We observed a very similar situation in our work on cesium
\cite{bouloufa2007}: we attribute this pathological behavior to an overly large number of
parameters, which
are thus strongly correlated.

In the second study on rubidium \cite{gutterres2002}, where the $rms$ value corresponds to a
very
satisfying $\chi^2$ value of 1.1, the evaluation of the uncertainty takes into account only
the binary
correlations between the parameters: the final error bar was therefore certainly
underestimated.

Concerning cesium, the uncertainty obtained in the first paper \cite{amiot2002} is one hundred
times smaller
than the one obtained in Ref \cite{bouloufa2007}, whereas the $rms$ value was notably smaller
in the
second paper. The bias introduced in the model by the omission of the two deepest levels can
partially
explain this somehow paradoxical situation. A slightly "wrong" model requires more adjustable
parameters,
with more restrained values, leading to worse agreement between the data. In addition, in Ref
\cite{amiot2002} like in Ref \cite{gutterres2002}, only binary correlations between the
parameters were
considered. In the second paper \cite{bouloufa2007}, we did a careful analysis of the
estimation of the
uncertainty of the lifetime value. What clearly appeared was that
the
interdependence of the parameters was very strong, probably because the number of parameters
was too
large. The $\chi^2$ value is rather small, which might be a clue to such a situation. We
tried to
linearize the model and to calculate the standard errors from the covariance matrix, as
described above:
the values obtained in this way were much too large for the linear approximation to remain
valid. We thus
tried to draw the contours of the $\chi^2$ function, but we found that the results of the
fitting
procedure depended in an unpredictable way on the allowed variation range of the parameters.
Like in Ref
\cite{freeland2001}, some free parameters did sometimes not move very much from their initial
value. The
consequence was that noticeably different parameter sets were yielding the same theory-experiment
agreement.
As available theoretical values of the long-range interaction coefficients are not precise
enough, it is
difficult to reduce the number of parameters of the fit. Reducing the experimental uncertainty
$\sigma_E$
should probably allow one to get rid of these difficulties and would certainly increase the
accuracy of
the lifetime determination.

The difficulties one might encounter in the evaluation of the uncertainty as the number of
parameters of
the model increases is certainly no reason to give up on the determination of atomic radiative lifetime
values
through PAS: the conclusion of this section is that a careful analysis of the errors coming
from the fit
itself must always be performed, and that such a study will guide one in the choice of the
model and of
the energy range of the levels introduced in the fit.

\section{Conclusion}
\label{sec:conclusion}

\begin{figure}[h]
\includegraphics[width=30pc]{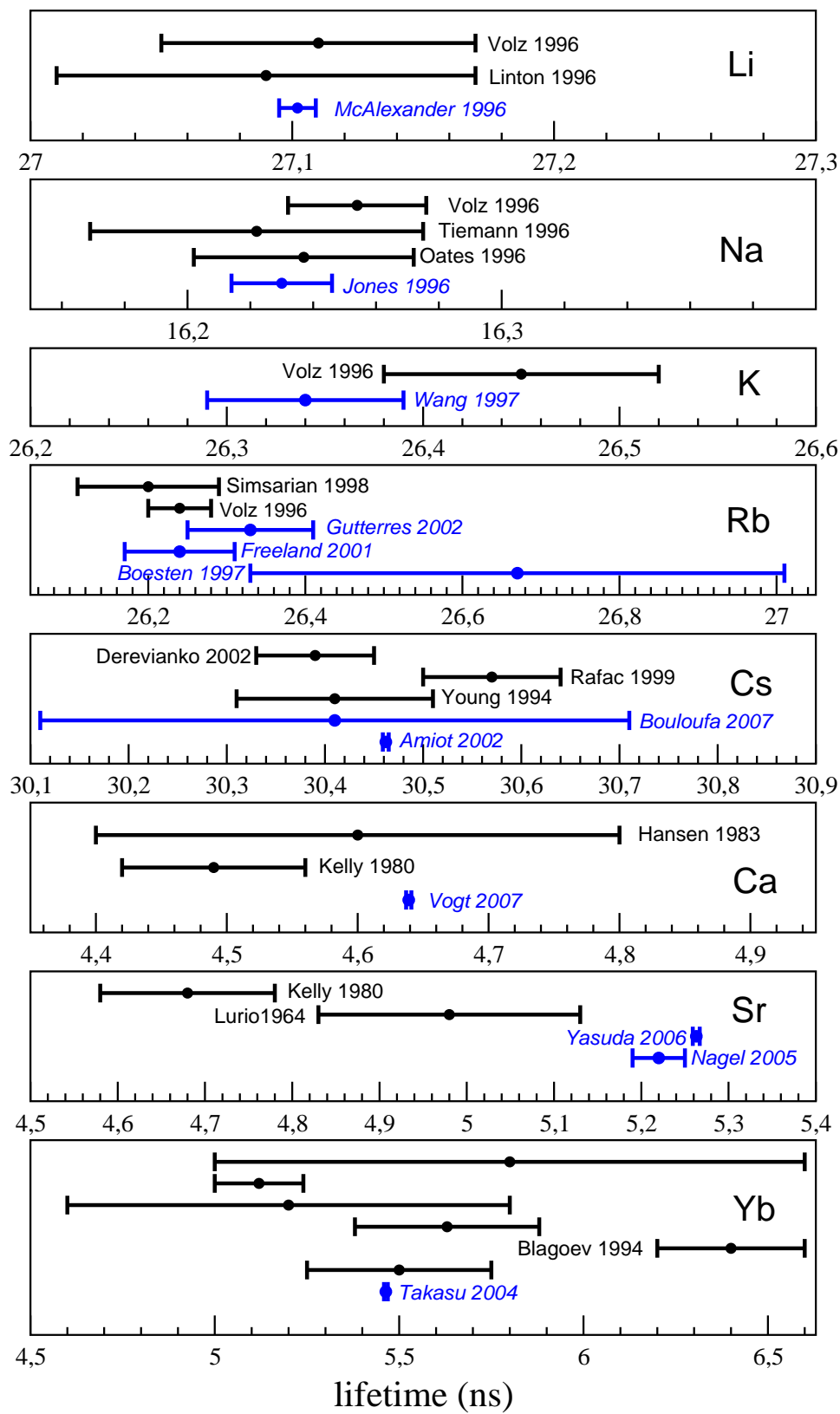}
\caption{\footnotesize Summary of the radiative lifetime values for all elements discussed in
the present
work. The names of authors using photoassociation spectroscopy are displayed in italics. For ytterbium,
all reported
values but the PAS one are extracted from the review of Blagoev and Komarovskii
\cite{blagoev1994}.}
\label{fig:lifetimes}
\end{figure}

In this paper we have reviewed recent experiments which derive atomic radiative lifetime
values from
photoassociative spectroscopy. This is possible because the energy spacing of the high-lying
molecular
states depends mainly on the long-range interatomic interaction, which itself depends on the
same atomic
radial integral as the atomic lifetime. Accurate values of atomic radiative lifetimes have
always been
difficult to obtain, both theoretically and experimentally. The emergence of the PAS method of
determination, based on a conceptually new approach, is thus very interesting. As potential
systematic
errors are quite different from those expected in atomic physics experiments, for instance, it
provides a
useful check on these lifetime determinations. The lifetime values discussed in this paper are
summarized
in Figure \ref{fig:lifetimes}, where the high quality of the PAS data is clearly visible.

The analysis consists in extracting, from the spectroscopic data, a precise long-range
coefficient: a
model is chosen for the calculation of the level energies and the parameters of the model are
fitted to
the experimental data. The model is either semi-classical or quantum-mechanical. In the first
case, a
two-parameter model, the so-called LeRoy-Bernstein law, was used in Sr and Yb studies.
Otherwise,
molecular potentials have to be considered. In principle, the long-range interaction has to be
connected
with {\it ab initio} or RKR potentials, as was the case for the Li, Ca and Sr
studies. In the
case of Na, K, Rb and Cs, the relevant potentials are called "pure long-range" and involve
only long-range
interaction parameters. The $R^{-3}$ behavior of the long-range interaction which is the basis
of the
lifetime determination is only asymptotically valid. A number of effects, which are likely to
contribute
to the interaction energy, have been introduced in the models, at the cost of an increased
number of
parameters.

Concerning the LeRoy-Bernstein approach, we first recall that its validity should always be
carefully
checked. Concerning the precision that can then be obtained, a very simple calculation allows
one to
predict the uncertainty on the lifetime value starting from the experimental value of the
spectroscopic
uncertainty and from the $v$ values of the vibrational levels introduced in the fit. The
predictions are
the same for quantum calculations as long as only two free parameters are needed. When
additional
parameters are introduced, the estimation of the uncertainty is more difficult. It has
probably been
sometimes underestimated, mainly because the role of the correlations between the
parameters was not
wholly considered. We recall a simple and general estimation of these effects, based on a
linearization of
the model. If it is not possible to apply it, contour calculations of the $\chi^2$ function
have to be
drawn and one has to carefully check the reproducibility of the convergence process: in the
cesium case,
for instance, we observed unpredictable results, due to an overly large number of parameters. We
hope to solve
this particular problem with the analysis of new experimental PAS results with improved
accuracy currently
in progress in our lab.

In spite of these difficulties, the ensemble of atomic radiative lifetimes results obtained
through PAS is
quite impressive and convincing. A number of accurate values have been derived, and they agree
well with
most of the previous results, obtained from accurate atomic physics measurements
\cite{volz1996}. The
agreement with available theoretical results is generally satisfying. It is even excellent in
the case of
lithium for which very precise calculations have been performed due to its simple atomic
structure.

Finally, the PAS method for extracting accurate atomic radiative lifetimes could represent a
promising
perspective for other elements which are nowadays laser-cooled and trapped, like Magnesium
\cite{sengstock1994}, Chromium \cite{bell1999}, Silver \cite{uhlenberg2000}, Erbium
\cite{berglund2007},
Francium \cite{sprouse1998}, and Radium \cite{guest2007}, provided that trapping densities are
high enough to
perform efficient PA experiments.

\section*{References}
\bibliographystyle{iopart-num}

\end{document}